\documentclass[epj,nopacs]{svjour}
\usepackage{graphics}
\begin{document}
\title{Correlator of heavy-quark currents at small \boldmath{$q^2$}\\
in the large-\boldmath{$\beta_0$} limit}
\author{A.G.~Grozin\thanks{grozin@particle.uni-karlsruhe.de}
\and C.~Sturm\thanks{sturm@particle.uni-karlsruhe.de}}
\institute{Institut f\"ur Theoretische Teilchenphysik, Universit\"at Karlsruhe, Germany}
\date{Received: December 2, 2004}
\abstract{The correlator of vector heavy-quark currents
at small $q^2$ is considered in the large-$\beta_0$ limit.
The leading IR renormalon ambiguity of the sum of the perturbative series
is canceled by the UV renormalon ambiguity of the gluon condensate.
Asymptotic behaviour of the perturbative series is obtained
in a model-independent way, up to a single unknown normalization factor.
Gluon-virtuality distribution functions for the perturbative correction
are calculated.}
\maketitle

\section{Introduction}
\label{Intro}

\begin{sloppypar}
Large-order behaviour of QCD perturbative series for various quantities
is being actively investigated in recent years~\cite{Be:99}.
Most of the results are obtained in the large-$\beta_0$ limit:
$\beta_0$ is considered a positive large parameter,
and series in $\beta_0\alpha_s\sim1$ are summed to all orders.
Strictly speaking, this can only happen in QCD
with $n_f\to-\infty$ flavours.
There is some empirical evidence that two-loop coefficients for many quantities
are well approximated by their leading $\beta_0$ terms
(naive nonabelianization~\cite{BG:95}).
This does not mean, however, that the large-order behaviour
of QCD perturbative series is reproduced by this limit.
It should be viewed as a toy model demonstrating possible patterns
of large-order behaviour of perturbative series,
and having some similarity to QCD.
\end{sloppypar}

The correlator of light-quark vector currents has been investigated,
at the first order in $1/\beta_0$, in the papers~\cite{PMP:84,Br:93}
(see also~\cite{Be:93}, where the result is presented as a double series,
not as a single series~\cite{Br:93}).
The result can also be re-written as an integral representation
with a transparent physical meaning~\cite{Ne:95}.
Techniques~\cite{PMP:84,Br:93} of calculations at the first order in $1/\beta_0$
are reviewed in~\cite{Gr:03}.

There is also a model-independent approach to the large-order behaviour
of QCD perturbative series based on renormalization group~\cite{Pa:78,BBK:97}.
It allows one to find the leading asymptotic behaviour,
as well as a few $1/L$ corrections to it
(where $L$ is the order of perturbation theory),
but leaves some normalization constants undetermined.
Some applications were considered in~\cite{Be:95,GN:97,CGM:03}.

In the present paper, we investigate the correlator of heavy-quark vector currents
at $q^2\ll m^2$, at the first order in $1/\beta_0$.
We demonstrate that the leading infrared (IR) renormalon ambiguity
in the perturbative part of this correlator is compensated
by the UV renormalon ambiguity in the gluon condensate,
which appears in the leading power correction (Sect.~\ref{CorrS}).
We apply the renormalization-group based method to obtain
the large-order behaviour of the perturbative series in a model-independent way,
up to a single unknown normalization factor (Sect.~\ref{Struct}).
We find the gluon-virtuality distribution functions
for the first two terms in the small-$q^2$ expansion of the correlator
(Sect.~\ref{Virt}).

\section{Correlator up to $1/\beta_0$}
\label{Corr}

\begin{figure}
\includegraphics{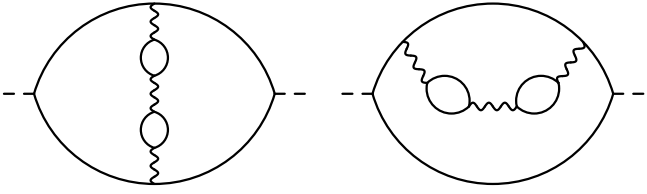}
\caption{Diagrams contributing to the correlator at $1/\beta_0$}
\label{Fig:Corr}
\end{figure}

We consider the correlator
\begin{equation}
i \int d x\,e^{iqx}
\left<T\{j_\mu(x),j_\nu(0)\}\right>
= (q_\mu q_\nu - q^2 g_{\mu\nu}) \Pi(q^2)
\label{Def}
\end{equation}
of two vector currents $j_\mu=\bar{Q}\gamma_\mu Q$,
and expand $\Pi(q^2)$ in $q^2/m^2$
($m$ is the mass of the heavy flavour $Q$).
To the first order in $1/\beta_0$,
the result can be obtained from diagrams like those in Fig.~\ref{Fig:Corr}.
The integrals in the coefficients of small-$q$ expansion
can be explicitly calculated in $\Gamma$-functions~\cite{Vl:80}.
The bare correlator can be written in the form
\begin{eqnarray}
&&\Pi(q^2) = \frac{N_c m_0^{-2\varepsilon}}{(4\pi)^{d/2}}
\sum_{n=0}^\infty \left(\frac{q^2}{m_0^2}\right)^n P_n \Gamma(n+\varepsilon)
\nonumber\\
&&{}\times \left[1 + \frac{1}{\beta_0} \sum_{L=1}^\infty
\frac{F_n(\varepsilon,L\varepsilon)}{L}
\left(\frac{\beta}{\varepsilon+\beta}\right)^L
+ \mathcal{O}\left(\frac{1}{\beta_0^2}\right)\right]\,,
\label{Form0}
\end{eqnarray}
where $N_c$ is the number of quark colours,
$m_0$ is the bare heavy-quark mass,
$d=4-2\varepsilon$ is the space-time dimension,
\begin{equation}
\beta = \beta_0 \frac{\alpha_s(\mu)}{4\pi}\,,\quad
\beta_0 = \frac{11}{3} C_A - \frac{4}{3} T_F n_f
\label{beta}
\end{equation}
is the $\beta$-function at the leading order in $1/\beta_0$
($n_f$ is the number of light flavours,
$T_F$ is the normalization factor
of the fundamental-representation generators $t^a$
defined by $\mathop{\mathrm{Tr}} t^a t^b = T_F \delta^{ab}$,
$C_F$ and $C_A$ are the Casimir operators
in the fundamental and the adjoint representations),
the leading-order coefficients are
\begin{equation}
P_0 = \frac{4}{3}\,,\quad
P_1 = \frac{4}{15}\,,\quad
P_2 = \frac{1}{35}\,,\quad\ldots
\label{Pn}
\end{equation}
the functions $F_n$ have the form
\begin{eqnarray}
&&F_n(\varepsilon,u) = C_F D(\varepsilon)^{u/\varepsilon-1} e^{\gamma\varepsilon}
\left(\frac{\mu^2}{m^2}\right)^u
\nonumber\\
&&{}\times \frac{\Gamma(1+u)\Gamma(1+n+u)\Gamma(2-u)\Gamma(n+u+\varepsilon)}%
{\Gamma(n+\varepsilon)\Gamma(3+n-\varepsilon)\Gamma(2+2n+2u)}
\nonumber\\
&&{}\times N_n(\varepsilon,u)
\label{Feu}
\end{eqnarray}
(where
\begin{equation}
D(\varepsilon) = 6 e^{\gamma\varepsilon} \Gamma(1+\varepsilon)
B(2-\varepsilon,2-\varepsilon) = 1 + \frac{5}{3} \varepsilon + \cdots
\label{De}
\end{equation}
comes from a light-quark loop,
$\gamma$ is the Euler constant),
and
\begin{eqnarray}
&&N_0(\varepsilon,u) = - 2
\Bigl[ 2\varepsilon (3-2\varepsilon)
+ (3+8\varepsilon-8\varepsilon^2) u
\nonumber\\
&&{}
+ (3-2\varepsilon) u^2 \Bigr]\,,
\nonumber\\
&&N_1(\varepsilon,u) = - \frac{4}{3}
\Bigl[ 18 (9-7\varepsilon^2+2\varepsilon^3)
\nonumber\\
&&{}
+ (265+60\varepsilon-319\varepsilon^2+96\varepsilon^3) u
\nonumber\\
&&{}
+ (150+16\varepsilon-161\varepsilon^2+48\varepsilon^3) u^2
\nonumber\\
&&{}
+ (35-19\varepsilon-4\varepsilon^2) u^3
+ (3-2\varepsilon) u^4 \Bigr]\,,
\nonumber\\
&&N_2(\varepsilon,u) = - \frac{2}{9}
\Bigl[ 1080 (72-54\varepsilon-11\varepsilon^2+13\varepsilon^3-2\varepsilon^4)
\nonumber\\
&&{}
+ 2 (63240-38917\varepsilon-28393\varepsilon^2+21620\varepsilon^3-3312\varepsilon^4) u
\nonumber\\
&&{}
+ (77878-33759\varepsilon-56299\varepsilon^2+35496\varepsilon^3-5184\varepsilon^4) u^2
\nonumber\\
&&{}
+ (23447-7432\varepsilon-17848\varepsilon^2+9366\varepsilon^3-1152\varepsilon^4) u^3
\nonumber\\
&&{}
+ (3895-1423\varepsilon-1627\varepsilon^2+478\varepsilon^3) u^4
\nonumber\\
&&{}
+ (417-190\varepsilon-68\varepsilon^2) u^5
+ 9 (3-2\varepsilon) u^6 \Bigr]\,,
\nonumber\\
&&\ldots
\label{Neu}
\end{eqnarray}

The functions $F_n(\varepsilon,u)$ at $n\ge1$ are regular at the origin,
and can be expanded in $\varepsilon$ and $u$.
Then the coefficient of $1/\varepsilon$ in the $1/\beta_0$ term
in the square bracket in~(\ref{Form0}) is~\cite{PMP:84}
\begin{equation}
- \frac{1}{2} \int_0^\beta \gamma_n(\beta) \frac{d\beta}{\beta}\,,
\label{Z1}
\end{equation}
where
\begin{equation}
\gamma_n(\beta) = - 2 \frac{\beta}{\beta_0} F_n(-\beta,0)\,.
\label{gamman}
\end{equation}
We should re-express the correlator~(\ref{Form0}) via the renormalized mass:
$m_0=Z_m m(\mu)$.
The $\overline{\mathrm{MS}}$ mass renormalization constant is,
up to the $1/\beta_0$ term,
\begin{equation}
Z_m(\beta) = 1
- \frac{1}{2\varepsilon} \int_0^\beta \gamma_m(\beta) \frac{d\beta}{\beta}
+ \frac{1}{2\varepsilon^2} \int_0^\beta \gamma_m(\beta)\,d\beta
+ \cdots
\label{Zm}
\end{equation}
where dots mean higher powers of $1/\varepsilon$,
and the mass anomalous dimension is~\cite{PMP:84}
\begin{equation}
\gamma_m(\beta) = 2 C_F \frac{\beta}{\beta_0}
\frac{1+(2/3)\beta}{B(2+\beta,2+\beta)\Gamma(3+\beta)\Gamma(1-\beta)}\,.
\label{gammam}
\end{equation}
Terms with $n\ge1$ in $\Pi(q^2)$~(\ref{Form0})
expressed via $m(\mu)$ and $\alpha_s(\mu)$ must be finite
at $\varepsilon\to0$.
In particular, the coefficient of $1/\varepsilon$ must vanish;
this happens if $\gamma_n(\beta)$~(\ref{gamman})
is related to $\gamma_m(\beta)$~(\ref{gammam}) by
\begin{equation}
\gamma_n(\beta) = 2 (n-\beta) \gamma_m(\beta)\,.
\label{gamma}
\end{equation}
This is indeed so, and not only for $n\ge1$, but also for $n=0$
(where this does not follow from the above argument).

We can also compare our result for $\Pi(0)$
with the correlator of massless-quark currents~\cite{PMP:84,Br:93}.
The $1/\beta_0$ term in $\beta_0 g_0^2 \Pi(0)$ can be rewritten as
\begin{equation}
\frac{4}{3} \frac{N_c}{\beta_0} \sum_{L=2}^\infty
\frac{F_0'(\varepsilon,L\varepsilon)}{L}
\left(\frac{\beta}{\varepsilon+\beta}\right)^L\,,
\label{Pi0}
\end{equation}
where
\begin{equation}
F_0'(\varepsilon,u) = \left(\frac{\mu^2}{m^2}\right)^\varepsilon e^{\gamma\varepsilon}
\Gamma(1+\varepsilon) \frac{u}{u-\varepsilon} F_0(\varepsilon,u-\varepsilon)\,.
\label{F0p}
\end{equation}
The coefficient of $1/\varepsilon$ in this sum is given by the formula
similar to~(\ref{Z1}), with the anomalous dimension
\begin{eqnarray}
&&\gamma(\beta) = - 2 \frac{\beta}{\beta_0} F_0'(-\beta,0)
\nonumber\\
&&{}= \frac{4}{3} \frac{C_F}{\beta_0}
\frac{\beta^2 (1-\beta) (1+2\beta)^2 (3+2\beta)^2 \Gamma(1+2\beta)}%
{(1+\beta)^2 (2+\beta) \Gamma(1-\beta) \Gamma^3(1+\beta)}\,.
\label{gamma0}
\end{eqnarray}
These ultraviolet divergences of the diagrams of Fig.~\ref{Fig:Corr}
don't feel the quark mass, and must be the same as in the massless case.
Inserting the colour factors into the QED result~\cite{Br:93},
we get the $1/\beta_0$ term in $\beta_0 g_0^2 \Pi(q^2)$ for massless quarks:
\begin{equation}
C_F N_c \frac{4\varepsilon}{\beta_0} \sum_{L=2}^\infty
\frac{F(\varepsilon,L\varepsilon)}{L}
\left(\frac{\beta}{\varepsilon+\beta}\right)^L\,,
\label{Piq2}
\end{equation}
where the function $F(\varepsilon,u)$ defined in~\cite{Br:93}
can be expressed via ${}_3F_2$ hypergeometric function of unit argument.
From the expression~(12) in~\cite{Br:93} for $F(\varepsilon,0)$,
we obtain the same anomalous dimension~(\ref{gamma0}).

\section{IR renormalons in the correlator}
\label{CorrS}

When expressed via $\alpha_s(\mu)$ and $m(\mu)$,
$\Pi(q^2)-\Pi(0)$ is finite at $\varepsilon\to0$:
\begin{equation}
\Pi(q^2)-\Pi(0) = \frac{N_c}{(4\pi)^2} \sum_{n=1}^\infty
\left(\frac{q^2}{m^2(\mu)}\right)^n P_n A_n(\mu)\,.
\label{Pimu}
\end{equation}
It does not depend on $\mu$;
substituting the solution of the renormalization-group equations,
\begin{eqnarray}
&&m(\mu) = \hat{m}
\left(\frac{\alpha_s(\mu)}{\alpha_s(\mu_0)}\right)^{\gamma_{m0}/(2\beta_0)}
K(\alpha_s(\mu))\,,
\nonumber\\
&&A_n(\mu) = \hat{A}_n
\left(\frac{\alpha_s(\mu)}{\alpha_s(\mu_0)}\right)^{n\gamma_{m0}/\beta_0}
K^{2n}(\alpha_s(\mu))\,,
\label{RGsol}
\end{eqnarray}
where $\gamma_{m0}=6C_F$ and
\begin{eqnarray}
&&K(\alpha_s) = \exp \int_0^{\alpha_s}
\left(\frac{\gamma_m(\alpha_s)}{2\beta(\alpha_s)}-\frac{\gamma_{m0}}{2\beta_0}\right)
\frac{d\alpha_s}{\alpha_s}
\nonumber\\
&&{} = 1 + \frac{\gamma_{m0}}{2\beta_0}
\left(\frac{\gamma_{m1}}{\gamma_{m0}} - \frac{\beta_1}{\beta_0}\right)
\frac{\alpha_s}{4\pi} + \cdots
\label{K}
\end{eqnarray}
we obtain the explicitly $\mu$-independent form
\begin{equation}
\Pi(q^2)-\Pi(0) = \frac{N_c}{(4\pi)^2} \sum_{n=1}^\infty
\left(\frac{q^2}{\hat{m}^2}\right)^n P_n \hat{A}_n\,.
\label{Pi1}
\end{equation}
At the $1/\beta_0$ order,
$\hat{A}_n$ can be expressed~\cite{Br:93} via $F_n(0,u)$:
\begin{equation}
\hat{A}_n = 1 + \frac{1}{\beta_0} \int_0^\infty S_n(u) e^{-u/\beta} du
+ \mathcal{O}\left(\frac{1}{\beta_0^2}\right)\,,
\label{Ahat}
\end{equation}
where $\beta$~(\ref{beta}) is taken at $\mu=\mu_0$, and
\begin{equation}
S_n(u) = \left.\frac{F_n(0,u)-F_n(0,0)}{u}\right|_{\mu=\mu_0}\,.
\label{Su}
\end{equation}
It is most convenient to use
\begin{equation}
\mu_0 = e^{-5/6} m
\label{mu0}
\end{equation}
(where $m$ is the on-shell mass), then
\begin{eqnarray}
&&S_n(u) = \frac{C_F}{(n-1)!\,(n+2)!}
\nonumber\\
&&{}\times \Biggl[
\frac{\Gamma(u)\Gamma(n+u)\Gamma(n+1+u)\Gamma(2-u)}{\Gamma(2n+2+2u)}
N_n(0,u)
\nonumber\\
&&{} - \frac{(n-1)!\,n!}{(2n+1)!} \frac{N_n(0,0)}{u} \Biggr]\,,
\label{Snu}
\end{eqnarray}
where
\begin{eqnarray}
&&N_0(0,u) = - 6 u (1+u)\,,
\nonumber\\
&&N_1(0,u) = - \frac{4}{3} (2+u) (81 + 92 u + 29 u^2 + 3 u^3)\,,
\nonumber\\
&&N_2(0,u) = - \frac{2}{9} (3+u)
(25920 + 33520 u + 14786 u^2
\nonumber\\
&&{} + 2887 u^3 + 336 u^4 + 27 u^5)\,,
\nonumber\\
&&\ldots
\label{Nu}
\end{eqnarray}
and
\begin{eqnarray}
&&N_0(0,0) = 0\,,\quad
N_1(0,0) = -216\,,
\nonumber\\
&&N_2(0,0) = -17280\,,\quad
\ldots
\label{N00}
\end{eqnarray}

Exact results for the correlator are known up to three loops~\cite{CKS:96};
for higher loops, we only know the terms with the largest power of $\beta_0$.
The perturbative series for $\hat{A}_{1,2}$ are presented in Appendix~\ref{Comp}.
The three-loop coefficients are not approximated well by the $\beta_0$ terms
(for $n_f=4$, 5).

The functions $S_n(u)$~(\ref{Snu}) have IR renormalon singularities
at $u=2$, 3, \dots{}
Therefore, the integral~(\ref{Ahat}) is not well-defined.
We can take the residue at the leading pole $u=2$
as a measure of the ambiguity of this integral:
\begin{equation}
\Delta \hat{A}_n = \frac{C_F}{\beta_0}
\frac{n(n+1)}{(2n+5)!} N_n(0,2) e^{10/3}
\left(\frac{\Lambda_{\overline{\mathrm{MS}}}}{m}\right)^4\,.
\label{DeltaA}
\end{equation}
UV renormalon singularities at $u=-1$, $-2$\dots{}
do not produce ambiguities in the sum of the perturbative series.

\begin{sloppypar}
The $T$-product of the currents in~(\ref{Def}) is given
by the operator product expansion
\begin{equation}
\Pi(q^2) = C_0(q^2) + \sum_i C_i(q^2) \frac{\left<O_i\right>}{m^{d_i}}\,,
\label{OPE}
\end{equation}
which separates short-distance contributions -- Wilson coefficients $C_i(q^2)$
and long-distance ones -- vacuum averages of operators $O_i$ with dimensions $d_i$.
Until now, we discussed the Wilson coefficient $C_0(q^2)$ of the unit operator,
or the perturbative part of $\Pi(q^2)$.
The leading power correction is given by the contribution
of the lowest-dimensional operator -- the vacuum condensate
$\left<\alpha_s G^a_{\mu\nu} G^a_{\mu\nu}\right>$
with dimension 4~\cite{SVZ:79}.
Namely, $\hat{A}_n$~(\ref{Ahat}) contains the power correction~\cite{SVZ:79}
\begin{equation}
- \frac{C_F}{6 N_g} \frac{n(n+1)(n+2)(n+3)}{2n+5}
\frac{\pi \left<\alpha_s G^a_{\mu\nu} G^a_{\mu\nu}\right>}{m^4}\,,
\label{GlueCond}
\end{equation}
where $N_g=C_F N_c/T_F$ is the number of gluon colours.
The gluon condensate has an UV renormalon~\cite{Za:92}.
We give a simple derivation of its UV renormalon ambiguity
in Appendix~\ref{App}.
The full correlator must be unambiguous:
the leading IR renormalon ambiguity of the leading Wilson coefficient $C_0(q^2)$
is canceled by the UV renormalon ambiguity of the vacuum condensate
in the leading power correction.
This explains why the leading IR renormalon is at $u=2$.
Combining~(\ref{DeltaA}), (\ref{GlueCond}), and~(\ref{DeltaG}),
we see that the renormalon ambiguities cancel if
\begin{equation}
N_n(0,2) = - \frac{3}{2} (n+2) (n+3) (2n+4)!\,.
\label{N02}
\end{equation}
And this is indeed so for $N_{1,2}(0,u)$~(\ref{Nu}).
\end{sloppypar}

Note that if we rewrite the correlator~(\ref{Pimu}), (\ref{Pi1})
via the on-shell mass $m$,
\begin{equation}
\Pi(q^2)-\Pi(0) = \frac{N_c}{(4\pi)^2} \sum_{n=1}^\infty
\left(\frac{q^2}{m^2}\right)^n P_n A_n\,,
\label{Pios}
\end{equation}
then
\begin{equation}
A_n = 1 + \frac{1}{\beta_0} \int_0^\infty \left[S_n(u) + 2n S_m(u)\right] e^{-u/\beta} du
+ \mathcal{O}\left(\frac{1}{\beta_0^2}\right)\,,
\label{Aos}
\end{equation}
where~\cite{BB:94}
\begin{equation}
S_m(u) = 6 C_F \left[ \frac{\Gamma(u)\Gamma(1-2u)}{\Gamma(3-u)} (1-u)
- \frac{1}{2u}\right]
\label{Sm}
\end{equation}
has the leading IR renormalon at $u=1/2$.
Therefore, the coefficients in the perturbative series for $A_n$
grow much faster than for $\hat{A}_n$.
This growth is dominated by the factor $(m/\hat{m})^{2n}$;
up to three loops, the exact perturbative coefficients in this factor,
and hence in $A_n=\hat{A}_n(m/\hat{m})^{2n}$,
are rather well approximated by the leading large-$\beta_0$ terms.

\section{Structure of the leading IR renormalon}
\label{Struct}

In this Section, we shall investigate
the coefficients of the perturbative series
\begin{equation}
\hat{A}_n = 1 + \sum_{L=1}^\infty c_{n,L}
\left(\frac{\alpha_s(\mu_0)}{4\pi}\right)^L
\label{Series}
\end{equation}
at $L\gg1$ using model-independent methods of~\cite{Pa:78,BBK:97}
(here $\mu_0$ is defined by~(\ref{mu0})).
The Borel images of these series are
\begin{eqnarray}
&&S_n(u) = \sum_{L=1}^\infty \frac{c_{n,L}}{(L-1)!}
\left(\frac{u}{\beta_0}\right)^{L-1}\,,
\label{SnuExact}\\
&&c_{n,L+1} = \left. \left(\beta_0 \frac{d}{du}\right)^L S_n(u) \right|_{u=0}\,.
\label{cnL}
\end{eqnarray}
We can formally invert this transformation:
\begin{equation}
\hat{A}_n = 1 + \frac{1}{\beta_0} \int_0^\infty S_n(u)
\exp\left[-\frac{4\pi}{\beta_0\alpha_s(\mu_0)} u\right]\,du\,;
\label{AhatExact}
\end{equation}
of course, this integral is ill-defined due to renormalon singularities at $u>0$.

These IR renormalon ambiguities ought to be compensated
by power corrections to the correlator.
Adding the gluon-condensate contribution~\cite{SVZ:79,BBIFTS:94} we require
\begin{eqnarray}
&&\hat{A}_n \left(\frac{q^2}{\hat{m}^2}\right)^n
- \frac{4}{3} a_n \frac{C_F}{N_g}
\frac{\pi \left<\alpha_s G^a_{\mu\nu} G^a_{\mu\nu}\right>_m}{m^4}
\nonumber\\
&&{}\times \left(1 + b_n \frac{\alpha_s(m)}{4\pi} + \cdots\right)
\left(\frac{q^2}{m^2}\right)^n
\label{Cond1}
\end{eqnarray}
to be free of leading ($\sim\Lambda_{\overline{\mathrm{MS}}}^4$)
renormalon ambiguities.
Here $a_1=3/7$, $a_2=5/3$ (see~(\ref{GlueCond})),
and the coefficients in the two-loop correction are~\cite{BBIFTS:94}
\begin{equation}
b_1 = \frac{135779}{3240}\,,\quad
b_2 = \frac{1969}{42}\,.
\label{bn}
\end{equation}
In fact, this correction has two colour structures $C_F$ and $C_A$;
unfortunately, only the results for $SU(3)_c$ are presented in~\cite{BBIFTS:94}.
We can re-express $\left<\alpha_s(\mu)(G^a_{\mu\nu} G^a_{\mu\nu})_\mu\right>$
via $\left<\beta(\alpha_s(\mu))(G^a_{\mu\nu} G^a_{\mu\nu})_\mu\right>$
which is $\mu$-independent to all orders.
Substituting also $\hat{m}/m$, we see that
\begin{equation}
\hat{A}_n - \frac{16\pi^2}{3} \frac{a_n}{\beta_0} \frac{C_F}{N_g}
\frac{\left<\beta G^a_{\mu\nu} G^a_{\mu\nu}\right>}{m^4}
\left(1 + \hat{b}_n \frac{\alpha_s(\mu_0)}{4\pi} + \cdots\right)\,,
\label{Cond2}
\end{equation}
where
\begin{equation}
\hat{b}_n = b_n - \frac{\beta_1}{\beta_0}
+ 2n \left[ C_F - \frac{\gamma_{m0}}{2\beta_0}
\left(\frac{\gamma_{m1}}{\gamma_{m0}} - \frac{\beta_1}{\beta_0}\right)
\right]\,,
\label{bhat}
\end{equation}
are free of leading renormalon ambiguities.

The UV renormalon ambiguity of the $\mu$-independent gluon condensate
$\left<\beta G^a_{\mu\nu} G^a_{\mu\nu}\right>$
can only be equal to $\Lambda_{\overline{\mathrm{MS}}}^4$
times a dimensionless constant.
We define
\begin{eqnarray}
&&\Delta \left<\beta G^a_{\mu\nu} G^a_{\mu\nu}\right> = N_G \Delta_0\,,
\nonumber\\
&&\Delta_0 = - \frac{3}{8\pi^2} N_g e^{10/3} \Lambda_{\overline{\mathrm{MS}}}^4
\label{DeltaExact}
\end{eqnarray}
(see~(\ref{DeltaG})).
In the large-$\beta_0$ limit, $N_G=1+\mathcal{O}(1/\beta_0)$;
in general, we cannot say much about this normalization factor,
except that it is some (unknown) constant of order 1.
Using
\begin{eqnarray}
&&\Lambda_{\overline{\mathrm{MS}}} = \mu_0
\exp\left[-\frac{2\pi}{\beta_0\alpha_s(\mu_0)}\right]
\left(\frac{\alpha_s(\mu_0)}{4\pi}\right)^{-\beta_1/(2\beta_0^2)}
\nonumber\\
&&\quad{}\times K_\beta(\alpha_s(\mu_0))\,,
\nonumber\\
&&K_\beta(\alpha_s) = \exp \int_0^{\alpha_s}
\left(\frac{1}{2\beta(\alpha_s)} - \frac{2\pi}{\beta_0\alpha_s}
+ \frac{\beta_1}{2\beta_0^2}\right) \frac{d\alpha_s}{\alpha_s}
\nonumber\\
&&{} = 1 + \frac{\beta_1^2-\beta_0\beta_2}{2\beta_0^3} \frac{\alpha_s}{4\pi}
+ \cdots
\label{Lambda}
\end{eqnarray}
we see that the leading IR renormalon ambiguity of $\hat{A}_n$
should be equal to $\exp\left[-8\pi/(\beta_0\alpha_s(\mu_0))\right]$
times some fractional powers of $\alpha_s(\mu_0)$.
In order to ensure this,
the Borel images should have a branching point at $u=2$:
\begin{equation}
S_n(u) = \sum_i \frac{r_i}{(2-u)^{1+a_i}} + \cdots
\label{Branching}
\end{equation}
where the dots mean a contribution regular at $u=2$.
Then
\begin{eqnarray}
\Delta \hat{A}_n &=& \frac{1}{\beta_0}
\exp\left[-\frac{8\pi}{\beta_0\alpha_s(\mu_0)}\right]
\nonumber\\
&&{}\times \sum_i \frac{r_i}{\Gamma(1+a_i)}
\left(\frac{\beta_0\alpha_s(\mu_0)}{4\pi}\right)^{-a_i}\,.
\label{DeltaAExact}
\end{eqnarray}

The requirement of the ambiguities cancellation
determines the behaviour of $S_n(u)$ at $u\to2$:
\begin{eqnarray}
&&S_n(u) = - \frac{2 a_n C_F N_G \Gamma(1+a) \beta_0^a}{(2-u)^{1+a}}
\nonumber\\
&&{}\times\left[1 + \frac{1}{\beta_0 a}
\left(\hat{b}_n + 2\frac{\beta_1^2-\beta_0\beta_2}{\beta_0^3}\right) (2-u)
+ \cdots\right]\,,
\label{SuExact}
\end{eqnarray}
where $a=2\beta_1/\beta_0^2$.
This result is model-independent; the power of $2-u$ is exact.
At the $1/\beta_0$ order, the formula~(\ref{SuExact}) reproduces the pole
of~(\ref{Snu}) at $u=2$.

The leading behaviour of $c_{n,L}$ at $L\gg1$ is determines,
according to~(\ref{cnL}), by the singularity of $S_n(u)$
nearest to the origin.
The leading contribution $\sim L! (-\beta_0)^L$
comes from the UV renormalon at $u=-1$.
It is sign-alternating, and thus not dangerous for the Borel summability.
The leading fixed-sign contribution comes from the IR renormalon at $u=2$.
Calculating multiple derivatives of~(\ref{SuExact}),
we obtain for this contribution
\begin{eqnarray}
&&c_{n,L+1} = - a_n C_F N_G L! \left(\frac{\beta_0}{2}\right)^L
\left(\frac{\beta_0 L}{2}\right)^a
\nonumber\\
&&{}\times \left[1 + \frac{2 b_n'}{\beta_0 L} + \cdots\right]\,,
\label{cLExact}
\end{eqnarray}
where
\begin{equation}
b_n' = \hat{b}_n + \frac{6\beta_1^2-4\beta_0\beta_2+\beta_0^2\beta_1}{2\beta_0^3}\,.
\label{bnp}
\end{equation}
The result~(\ref{cLExact}) is also model-independent.
It contains a single unknown normalization constant $N_G$ (common to all $n$).
The power $a$ of $L$ in~(\ref{cLExact}) is exact.
The coefficients $b_n'$ are, for $SU(3)_c$,
\begin{eqnarray}
&&b_1' = \frac{4289}{3240} + \frac{5905}{6} \frac{1}{\beta_0}
- \frac{119531}{12} \frac{1}{\beta_0^2} + \frac{34347}{\beta_0^3}\,,
\nonumber\\
&&b_2' = \frac{193}{84} + \frac{6449}{6} \frac{1}{\beta_0}
- \frac{129803}{12} \frac{1}{\beta_0^2} + \frac{34347}{\beta_0^3}\,.
\label{bn3}
\end{eqnarray}
For $n_f=5$, for example, the full coefficient of $1/L$ in~(\ref{cLExact})
is of order 10 (for both $n=1$ and 2);
this means that this asymptotics is only applicable for $L\gg10$.

\section{Virtuality distribution functions}
\label{Virt}

\begin{sloppypar}
The renormalization-group invariants~(\ref{Ahat}) at the $1/\beta_0$ order
can be rewritten in the form of the leading (one-gluon) perturbative correction,
but with the running (one-loop) $\alpha_s$ under the integral sign~\cite{Ne:95}:
\begin{equation}
\hat{A}_n = 1 + \int_0^\infty w_n(\tau)
\frac{\alpha_s(\sqrt{\tau}\mu_0)}{4\pi} \frac{d\tau}{\tau}
+ \mathcal{O}\left(\frac{1}{\beta_0^2}\right)\,,
\label{Neubert}
\end{equation}
where gluon-virtuality distribution functions $w_i(\tau)$
are given by
\begin{equation}
w_n(\tau) = \frac{1}{2 \pi i} \int_{-i\infty}^{+i\infty}
S_n(u) \tau^u du\,.
\label{wt}
\end{equation}
If $\tau<1$, we can close the contour to the right,
and find $w_i(\tau)$ as the sum of residues of the poles at $u>0$
(IR renormalons).
If $\tau>1$, the sum of the residues of UV renormalons ($u<0$)
is calculated instead.
\end{sloppypar}

\begin{figure}
\begin{picture}(87,63)
\put(45.5,31.5){\makebox(0,0){\includegraphics{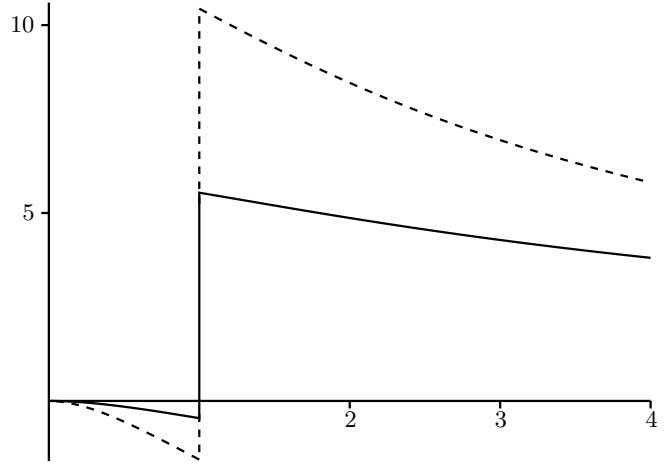}}}
\put(46,8){\makebox(0,0)[t]{2}}
\put(66,8){\makebox(0,0)[t]{3}}
\put(86,8){\makebox(0,0)[t]{4}}
\put(4,34.5){\makebox(0,0)[r]{5}}
\put(4,59.5){\makebox(0,0)[r]{10}}
\end{picture}
\caption{Functions $w_1(\tau)/C_F$ (solid line) and  $w_2(\tau)/C_F$ (dashed line)}
\label{Fig:w12}
\end{figure}

Our result is
\begin{eqnarray}
&&w_1(\tau) = C_F \Biggl\{ \frac{2}{9} \frac{1}{\tau(4+\tau)^4}
\biggl[ - 60 \Bigl(80 - 312 \tau - 368 \tau^2
\nonumber\\
&&{} - 122 \tau^3 - 18 \tau^4 - \tau^5\Bigr)
\frac{1}{\sqrt{\tau(4+\tau)}} \log\frac{\sqrt{\tau}+\sqrt{4+\tau}}{2}
\nonumber\\
&&{}+ 1200 - 4880 \tau - 4700 \tau^2 - 2062 \tau^3 - 391 \tau^4 - 27 \tau^5
\biggr]
\nonumber\\
&&{} + 6 \theta(\tau-1) \Biggr\}\,,
\nonumber\\
&&w_2(\tau) = C_F \Biggl\{ \frac{1}{27} \frac{1}{\tau^2(4+\tau)^6}
\biggl[ 840 \Bigl(1512 - 564 \tau + 356 \tau^2
\nonumber\\
&&{} + 2290 \tau^3 + 1196 \tau^4 + 274 \tau^5 + 26 \tau^6 + \tau^7\Bigr)
\nonumber\\
&&{}\times\frac{1}{\sqrt{\tau(4+\tau)}} \log\frac{\sqrt{\tau}+\sqrt{4+\tau}}{2}
\nonumber\\
&&{} - 317520 + 171360 \tau - 105084 \tau^2 - 462224 \tau^3
\nonumber\\
&&{} - 543492 \tau^4
- 254664 \tau^5 - 60209 \tau^6 - 7008 \tau^7 - 324 \tau^8
\biggr]
\nonumber\\
&&{} + 12 \theta(\tau-1) \Biggr\}\,.
\label{w12}
\end{eqnarray}
These functions are shown in Fig.~\ref{Fig:w12}.

The functions $S_n(u)$~(\ref{Snu}) are regular at $u=0$:
the singularity of the first term is canceled by the second one.
We may retain only the first term in~(\ref{Snu}) when calculating $w_n(\tau)$,
and use the contour to the right of the origin if $\tau<1$
and to the left of it if $\tau>1$.
This first term falls off fast when $\mathop{\mathrm{Im}}u\to\pm\infty$,
and we may close the contour in any way we like.
Therefore, the expressions for $w_n(\tau)$ for $\tau>1$
differ from those for $\tau<1$
by the residue of the first term at $u=0$.
In other words, the distribution functions contain $\theta(\tau-1)$
with the coefficient
\begin{equation}
-F_n(0,0) = - C_F \frac{N_n(0,0)}{(n+1)(n+2)(2n+1)!}\,,
\label{ThetaCoeff}
\end{equation}
as confirmed by~(\ref{w12}).
Such $\theta(\tau-1)$ terms appear in distribution functions
when the leading (one-gluon) perturbative correction diverges,
i.e., $F(0,0)\ne0$, see, e.g., \cite{Ne:95,Gr:03,GN:97}.

The behaviour of $w_n(\tau)$ at $\tau\to0$ is determined
by the leading IR renormalon.
It is at $u=2$, and hence $w_n(\tau)\sim\tau^2$:
\begin{equation}
w_1(\tau) = - \frac{6}{7} C_F \tau^2 + \cdots\quad
w_2(\tau) = - \frac{10}{3} C_F \tau^2 + \cdots
\label{wt0}
\end{equation}
At $\tau\to\infty$, it is determined by the leading UV renormalon
at $u=-1$:
\begin{eqnarray}
&&w_1(\tau) = \frac{2}{3} C_F \left(10 \log\tau + \frac{41}{3}\right) \frac{1}{\tau}
+ \cdots
\nonumber\\
&&w_2(\tau) = \frac{256}{9} C_F \frac{1}{\tau} + \cdots
\label{wtinf}
\end{eqnarray}
UV renormalon singularities in $S_n(u)$ are double poles,
and hence $w_n(\tau)$ behave not just as $1/\tau$,
but contain logarithmic terms.

\begin{sloppypar}
The integrals~(\ref{Neubert}) are ill-defined, just like~(\ref{Ahat}).
The one-loop running $\alpha_s(\sqrt{\tau}\mu_0)$ has a pole
at small $\tau=(e^{5/6}\Lambda_{\overline{\mathrm{MS}}}/m)^4$,
and we integrate across it.
The ambiguity is given by the residue;
using the small-$\tau$ asymptotics~(\ref{wt0}),
we reproduce~(\ref{DeltaA}).
\end{sloppypar}

\section{Conclusion}
\label{Conc}

In this paper, we have considered the correlator of two heavy-quark vector currents,
expanded up to $\mathcal{O}(q^4)$, at the first order in $1/\beta_0$.
The $d$-dimensional bare result is given by~(\ref{Form0})--(\ref{Neu}).
All $1/\varepsilon$ divergences cancel in $\Pi(q^2)-\Pi(0)$
expressed via the renormalized coupling and mass.
The finite part is given by~(\ref{Pi1})--(\ref{N00}).
This result contains highest powers of $\beta_0$ in all orders of perturbation theory,
and can be used for checking future multiloop calculations.
The Borel image $S_n(u)$ has IR renormalon singularities at $u=2$, 3\dots{}
making the series not Borel-summable.
The corresponding ambiguities in the sum of the perturbative series
for the correlator are compensated by the UV renormalon ambiguities
of the vacuum condensates which appear in power corrections.
This is explicitly demonstrated for the leading renormalon ($u=2$)
and the leading power correction (gluon condensate),
at the order $1/\beta_0$.

The structure of this leading IR renormalon can be studied
in a model-independent way, beyond the large-$\beta_0$ limit,
on the basis of the renormalization-group properties of the gluon condensate.
This singularity is a branching point~(\ref{SuExact}),
where the power of $2-u$ is known exactly,
and $N_G$ is an unknown normalization factor
(one constant for all $S_n(u)$).
The leading fixed-sign asymptotics of the perturbative coefficients~(\ref{cLExact})
is also a model-independent QCD result
(we don't consider a larger alternating-sign contribution
of the UV renormalon at $u=-1$, which is not dangerous for Borel summability).

The $1/\beta_0$-order results can be rewritten in the form of the leading
(one-gluon) perturbative correction, but with the (1-loop) running $\alpha_s$
under the integral sign.
The functions in these integrals have the meaning of the distribution functions
in the gluon virtuality in the one-gluon corrections.
They are presented in~(\ref{w12}), Fig.~\ref{Fig:w12}.
Their behaviour at small virtualities is determined by the leading IR renormalon;
in our case, it is at $u=2$, and contributions from small virtualities are
strongly suppressed (as $\tau^2$).
The behaviour at large virtualities is determined by the leading UV renormalon
at $u=-1$, and the contributions of large virtualities fall off slowly,
as $1/\tau$ (up to logarithms).

\begin{acknowledgement}
We are grateful to K.G.~Chetyrkin for his great help and numerous discussions.
A.G.G.'s work was supported by SFB/TR 9 (Computetional Particle Physics).
C.S. would like to thank the Graduiertenkolleg
\emph{``Hochenergiephysik und Teilchenastrophysik''} for financial support.
\end{acknowledgement}

\appendix
\section{UV renormalon in the gluon condensate}
\label{App}

\begin{figure}
\includegraphics{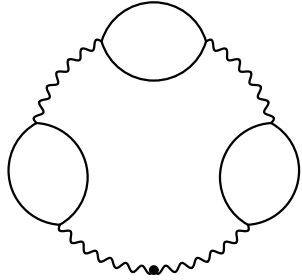}
\caption{Perturbative contribution to the gluon condensate}
\label{Fig:Glue}
\end{figure}

Let's consider the perturbative contribution to the vacuum average
of the bare operator $g_0^2 G^a_{0\mu\nu} G^a_{0\mu\nu}$
with a sharp infrared cutoff $\lambda$
(which should be large enough, so that the perturbation theory makes sense).
This operator is not renormalized at one loop:
$g_0^2 G^a_{0\mu\nu} G^a_{0\mu\nu}=4\pi\alpha_s(\mu)(G^a_{\mu\nu}G^a_{\mu\nu})_\mu$.
At the order $1/\beta_0$, we need to calculate diagrams in Fig.~\ref{Fig:Glue}.
The result can be written as
\begin{equation}
4\pi \left<\alpha_s G^a_{\mu\nu} G^a_{\mu\nu}\right>_\lambda
= \frac{1}{\beta_0} \sum_{L=1}^\infty
\frac{F(\varepsilon,L\varepsilon)}{L}
\left(\frac{\beta}{\varepsilon+\beta}\right)^L
+ \mathcal{O}\left(\frac{1}{\beta_0^2}\right)\,,
\label{GForm}
\end{equation}
where
\begin{equation}
F(\varepsilon,u) = - 2 N_g D(\varepsilon)^{u/\varepsilon-1} e^{\gamma\varepsilon}
\left(\frac{\mu^2}{\lambda^2}\right)^u \lambda^4
\frac{3-2\varepsilon}{\Gamma(2-\varepsilon)} \frac{u}{2-u}\,,
\label{GFeu}
\end{equation}
$N_g=C_F N_c/T_F$ is the number of gluon colours.
The coefficient of $1/\varepsilon$ vanishes in~(\ref{GForm}),
as expected, because $F(\varepsilon,0)=0$.
Therefore, it is finite at $\varepsilon\to0$:
\begin{equation}
4\pi \left<\alpha_s G^a_{\mu\nu} G^a_{\mu\nu}\right>_\lambda
= \frac{1}{\beta_0} \int_0^\infty S(u) e^{-u/\beta} du
+ \mathcal{O}\left(\frac{1}{\beta_0^2}\right)\,,
\label{Ghat}
\end{equation}
where
\begin{equation}
S(u) = \frac{F(0,u)-F(0,0)}{u}
= - \frac{6 N_g \lambda^4}{2-u}
\label{GSu}
\end{equation}
if we choose $\mu_0=e^{-5/6}\lambda$.
The gluon condensate has UV renormalon at $u=2$;
its ambiguity is given by the residue:
\begin{equation}
\Delta \pi \left<\alpha_s G^a_{\mu\nu} G^a_{\mu\nu}\right>
= - \frac{3}{2} \frac{N_g}{\beta_0} e^{10/3}
\Lambda_{\overline{\mathrm{MS}}}^4\,.
\label{DeltaG}
\end{equation}

\section{Perturbative series for $\hat{A}_{1,2}$}
\label{Comp}

\begin{eqnarray*}
&&\hat{A}_1 = 1
+ \Biggl\{ \frac{329}{27}
+ \frac{15C_F+16C_A}{\beta_0}\\
&&\quad{} - \frac{6C_A(11C_F+7C_A)}{\beta_0^2} \Biggr\}
C_F \frac{\alpha_s(\mu_0)}{4\pi}\\
&&{} + \Biggl\{ - \frac{1639}{81} \beta_0
+ \left( \frac{120817}{576} \zeta_3 - \frac{63323}{288} \right) C_F\\
&&\quad{}+ \left( \frac{4183}{128} \zeta_3 - \frac{170077}{5184} \right) C_A
+ \left( \frac{1015}{72} \zeta_3 - \frac{25187}{972} \right) T_F\\
&&\quad{} + \biggl[ \frac{1643}{18} C_F^2
+ \left( - 132 \zeta_3 + \frac{13795}{54} \right) C_F C_A\\
&&\qquad{} + \left( 132 \zeta_3 - \frac{131}{6} \right) C_A^2
\biggr] \frac{1}{\beta_0}\\
&&\quad{} + \biggl( \frac{225}{2} C_F^3 - \frac{3296}{9} C_F^2 C_A
- \frac{1661}{18} C_F C_A^2 + \frac{161}{8} C_A^3 \biggr) \frac{1}{\beta_0^2}\\
&&\quad{} - \frac{3 C_A (11 C_F + 7 C_A) (30 C_F^2 + 43 C_F C_A + 7 C_A^2)}{\beta_0^3}\\
&&\quad{} + \frac{18 C_F C_A^2 (11 C_F + 7 C_A)^2}{\beta_0^4} \Biggr\}
C_F \left(\frac{\alpha_s(\mu_0)}{4\pi}\right)^2\\
&&{} + \left\{ \left( - 24 \zeta_3 + \frac{20398}{243} \right) \beta_0^2 + \cdots \right\}
C_F \left(\frac{\alpha_s(\mu_0)}{4\pi}\right)^3\\
&&{} + \left\{ \left( 108 \zeta_4 + \frac{1316}{9} \zeta_3 - \frac{122120}{243} \right)
\beta_0^3 + \cdots \right\}\\
&&\quad{}\times C_F \left(\frac{\alpha_s(\mu_0)}{4\pi}\right)^4 + \cdots\\
&&\hat{A}_2 = 1
+ \Biggl\{ \frac{2333}{135}
+ \frac{2(15C_F+16C_A)}{\beta_0}\\
&&\quad{} - \frac{12 C_A (11 C_F + 7 C_A)}{\beta_0^2} \Biggr\}
C_F \frac{\alpha_s(\mu_0)}{4\pi}\\
&&{} + \Biggl\{ - \frac{392279}{16200} \beta_0
+ \left( \frac{34224293}{9216} \zeta_3 - \frac{8150760227}{1866240} \right) C_F\\
&&\quad{} + \left( \frac{22668817}{55296} \zeta_3 - \frac{1822035101}{3732480} \right) C_A\\
&&\quad{} + \left( \frac{497105}{18432} \zeta_3 - \frac{353936273}{6220800} \right) T_F\\
&&\quad{} + \biggl[ \frac{3019}{9} C_F^2
+ \left( - 264 \zeta_3 + \frac{90991}{135} \right) C_F C_A\\
&&\qquad{} + \left( 264 \zeta_3 - \frac{131}{3} \right) C_A^2
\biggr] \frac{1}{\beta_0}\\
&&\quad{} + \biggl( 450 C_F^3 - \frac{41632}{45} C_F^2 C_A
- \frac{16049}{45} C_F C_A^2 + \frac{161}{4} C_A^3 \biggr) \frac{1}{\beta_0^2}\\
&&\quad{} - \frac{6 C_A (11 C_F + 7 C_A) (60 C_F^2 + 75 C_F C_A + 7 C_A^2)}{\beta_0^3}\\
&&\quad{} + \frac{72 C_F C_A^2 (11 C_F + 7 C_A)^2}{\beta_0^4} \Biggr\}
C_F \left(\frac{\alpha_s(\mu_0)}{4\pi}\right)^2\\
&&{} + \left\{ \left( - 48 \zeta_3 + \frac{26770423}{243000} \right) \beta_0^2 + \cdots \right\}
C_F \left(\frac{\alpha_s(\mu_0)}{4\pi}\right)^3\\
&&{} + \left\{ \left( 216 \zeta_4 + \frac{9332}{45} \zeta_3 - \frac{791737663}{1215000} \right)
\beta_0^3 + \cdots \right\}\\
&&\quad{}\times C_F \left(\frac{\alpha_s(\mu_0)}{4\pi}\right)^4 + \cdots
\end{eqnarray*}
Numerically,
\begin{eqnarray*}
&&\hat{A}_1 = 1
+ \left( 4.06173 + \frac{22.6667}{\beta_0} - \frac{214}{\beta_0^2} \right)
\frac{\alpha_s(\mu_0)}{\pi}\\
&&{} + \biggl( - 1.68621 \beta_0 + 4.82974 + \frac{148.415}{\beta_0}
- \frac{187.540}{\beta_0^2}\\
&&\quad{} - \frac{7712.92}{\beta_0^3} + \frac{22898}{\beta_0^4}
\biggr) \left(\frac{\alpha_s(\mu_0)}{\pi}\right)^2\\
&&{} + (1.14777 \beta_0^2 + \cdots) \left(\frac{\alpha_s(\mu_0)}{\pi}\right)^3\\
&&{} + (-1.09319 \beta_0^3 + \cdots) \left(\frac{\alpha_s(\mu_0)}{\pi}\right)^4
+ \cdots\\
&&\hat{A}_2 = 1
+ \left( 5.76049 + \frac{45.3333}{\beta_0} - \frac{428}{\beta_0^2} \right)
\frac{\alpha_s(\mu_0)}{\pi}\\
&&{} + \biggl( - 2.01790 \beta_0 + 10.8546 + \frac{373.841}{\beta_0}
- \frac{588.373}{\beta_0^2}\\
&&\quad{} - \frac{25127.2}{\beta_0^3} + \frac{91592}{\beta_0^4}
\biggr) \left(\frac{\alpha_s(\mu_0)}{\pi}\right)^2\\
&&{} + (1.09307 \beta_0^2 + \cdots) \left(\frac{\alpha_s(\mu_0)}{\pi}\right)^3\\
&&{} + (-0.87799 \beta_0^3 + \cdots) \left(\frac{\alpha_s(\mu_0)}{\pi}\right)^4
+ \cdots
\end{eqnarray*}

\end{document}